\documentclass[conference]{IEEEtran}
\IEEEoverridecommandlockouts
\usepackage{cite}
\usepackage{amsmath,amssymb,amsfonts}
\usepackage{array} % For custom column types
\usepackage{algorithmic}
\usepackage{graphicx}
\usepackage{hyperref} 
\usepackage{newtxtext,newtxmath}
\usepackage{textcomp}
\usepackage{booktabs}
\usepackage{amsmath}
\usepackage{algorithm}
\usepackage{xcolor}
\def\BibTeX{{\rm B\kern-.05em{\sc i\kern-.025em b}\kern-.08em
    T\kern-.1667em\lower.7ex\hbox{E}\kern-.125emX}}
\IEEEoverridecommandlockouts\IEEEpubid{\makebox[\columnwidth]{979-8-3315-0376-5/25/\$31.00 $\copyright$2025 IEEE \hfill}\hspace{\columnsep}\makebox[\columnwidth]{ }}
\begin{document}

\title{Mamba4Net: Distilled Hybrid Mamba Large Language Models For Networking}

\author{\IEEEauthorblockN{Linhan Xia}
\IEEEauthorblockA{\textit{ICNLab, Shenzhen Graduate School,} \\
\textit{Peking University, }\\
Shenzhen, P.R.China \\
linhan.xia@ou.edu}
\and
\IEEEauthorblockN{Mingzhan Yang}
\IEEEauthorblockA{\textit{ICNLab, Shenzhen Graduate School,} \\
\textit{Peking University, }\\
Shenzhen, P.R.China \\
my47@illinois.edu}
\and
\IEEEauthorblockN{Jingjing Wang}
\IEEEauthorblockA{\textit{ICNLab, Shenzhen Graduate School,} \\
\textit{Peking University, }\\
Shenzhen, P.R.China \\
jingjing\_wang@pku.edu.cn}
\and
\IEEEauthorblockN{Ziwei Yan}
\IEEEauthorblockA{\textit{ICNLab, Shenzhen Graduate School,} \\
\textit{Peking University, }\\
Shenzhen, P.R.China \\
yanzw@pku.edu.cn}
\and
\IEEEauthorblockN{Yakun Ren}
\IEEEauthorblockA{\textit{Sf Technology,} \\
Shenzhen, P.R.China \\
yakunren@sfexpress.com}
\and
\IEEEauthorblockN{Guo Yu}
\IEEEauthorblockA{\textit{GienTech Technology Co.,Ltd.,} \\
Shenzhen, P.R.China \\
yu.guo18@gientech.com}
\and
\IEEEauthorblockN{Kai Lei*}
\IEEEauthorblockA{\textit{ICNLab, Shenzhen Graduate School,} \\
\textit{Peking University, }\\
Shenzhen, P.R.China \\
leik@pkusz.edu.cn\\
Corresponding author}
}

\maketitle
\begin{abstract}
Transformer-based large language models (LLMs) are increasingly being adopted in networking research to address domain-specific challenges. However, their quadratic time complexity and substantial model sizes often result in significant computational overhead and memory constraints, particularly in resource-constrained environments. Drawing inspiration from the efficiency and performance of the Deepseek-R1 model within the knowledge distillation paradigm, this paper introduces Mamba4Net, a novel cross-architecture distillation framework. Mamba4Net transfers networking-specific knowledge from transformer-based LLMs to student models built on the Mamba architecture, which features linear time complexity. This design substantially enhances computational efficiency compared to the quadratic complexity of transformer-based models, while the reduced model size further minimizes computational demands, improving overall performance and resource utilization. To evaluate its effectiveness, Mamba4Net was tested across three diverse networking tasks: viewport prediction, adaptive bitrate streaming, and cluster job scheduling. Compared to existing methods that do not leverage LLMs, Mamba4Net demonstrates superior task performance. Furthermore, relative to direct applications of transformer-based LLMs, it achieves significant efficiency gains, including a throughput 3.96 times higher and a storage footprint of only 5.48\% of that required by previous LLM-based approaches. These results highlight Mamba4Net’s potential to enable the cost-effective application of LLM-derived knowledge in networking contexts. The source code is openly available to support further research and development.
\end{abstract}


\section{Introduction}
As global data traffic continues to surge exponentially, optimizing and controlling networking systems have become imperative to ensure seamless connectivity and performance. Deep learning technologies are at the forefront of this optimization, being employed to address a variety of networking tasks such as bandwidth prediction \cite{bentaleb2022bob}, congestion control \cite{donta2023icocoa} \cite{huang2024copa}, and traffic analysis \cite{shen2022machine}\cite{lin2022bert}. These advanced algorithms enable more efficient networking management and resource allocation, contributing to improved networking reliability and user experience. However, traditional deep learning models present significant challenges in this domain. The adaptation of these models to different tasks often necessitates high engineering costs, as each new application may require bespoke model architectures and extensive retraining. Furthermore, these models frequently exhibit suboptimal performance when deployed in environments with unseen data distributions, thereby limiting their robustness and generalizability \cite{zhang2021sensei}.These limitations impede the scalability and practicality of implementing deep learning solutions across diverse and dynamic network environments.

To address these limitations, recent research has explored the application of Transformer-based Large Language Models (LLMs) in networking, giving rise to LLMs for Networks (LLMs4Net) \cite{wu2024netllm}. These approaches leverage the pre-trained, network-specific knowledge of LLMs to solve multiple network problems with a single model, significantly reducing engineering costs and demonstrating promising performance. 

Nevertheless, existing LLMs4Net approaches are constrained by two primary limitations:

\begin{itemize}
    \item \textbf{Excessive Computational Overhead.} The Transformer backbone of LLMs operates with quadratic time complexity, making it unsuitable for real-world network applications\cite{keles2023computational} where data flows are large and continuous. This high computational demand poses significant challenges for deployment on resource-constrained network devices such as routers and AR headsets, which lack the necessary processing power to handle such intensive workloads.

    \item \textbf{Significant Redundant Storage Consumption.} 
    LLMs encompass a vast array of knowledge beyond networking, leading to substantial memory usage in the form of numerous parameters \cite{kwon2023efficient}. This extraneous information not only fails to contribute to networking-specific tasks but also escalates the operational costs associated with deploying these models in real-world scenarios.
\end{itemize}

%这一段讲一下我们研究的philosophy。
To overcome the limitations identified in prior LLMs4Net approaches, this study proposes a novel framework for extracting domain-specific knowledge tailored to networking applications from LLMs. By focusing on networking-relevant information, this approach mitigates the influence of extraneous knowledge, resulting in a more compact model size. Furthermore, it significantly reduces time complexity, thereby lowering both training and inference overheads. Consequently, this method enhances the cost-effectiveness of deploying large language models within the networking domain. Inspired by the impressive breakthrough of Deepseek-R1 model \cite{guo2025deepseek}, which employed knowledge distillation to achieve impressive performance and low training cost, we develop \textbf{Mamba4Net}. Mamba4Net is a framework that distills the networking-specific knowledge of large Transformer-based LLMs into a more compact hybrid Mamba-based model. This approach enables the compression of the model’s scale, leading to a reduction in memory usage and an enhancement in computational efficiency. Specifically, it transforms the complexity from quadratic to linear, thereby minimizing computational overhead. However, previous knowledge distillation approaches suffer from unstable and inefficient training processes due to random weight initialization.

To tackle these challenges and achieve the extraction of networking-specific knowledge and effective cross-architecture distillation, we make the following contributions:
\begin{enumerate}
    
    \item \textbf{Domain Knowledge-Oriented Cross-Heterogeneous Distillation (DKO):} 
A novel method that distills networking-specific knowledge from Transformer LLMs to Mamba models, reducing quadratic complexity to linear while filtering out irrelevant information.

    \item \textbf{Cross-Heterogeneous Weight Reusing (denoted as CWR):} We propose a creative weights initialization method, namely cross-heterogeneous weight reusing, which has been developed to minimize the instability and computational overhead during the cross-heterogeneous distillation phase. This approach differs from the traditional method of randomly initializing the student model by transforming the teacher's attention block weights (from Transformer-based LLMs) into compact factors by decomposing high-dimensional matrices into low-rank subspaces. These compact factors are then used to initialize the Mamba-based student's parameters, effectively narrowing the representation gap and facilitating more effective knowledge transfer. By focusing on the essential components of the teacher model's weights, CWR avoids unnecessary overhead and provides a robust, architecture-aware method for achieving efficient distillation across disparate neural network designs. 
    \item \textbf{Comprehensive Experiments and Open-Sourced Code:} We conducted a series of experiments to validate the performance and cost of the proposed \textbf{Mamba4Net}, which has been employed in three distinct networking tasks, including viewport prediction, adaptive bitrate streaming, and cluster job scheduling (the same as for NetLLM). The experiments results demonstrate consistent improvements in both performance and computational efficiency, particularly in resource-constrained scenarios where resources are limited. The experimental code is available on GitHub, ensuring reproducibility and enabling further exploration by the research community.
\end{enumerate}

%插入三个任务
Our work take three networking tasks at use cases.

The remainder of this paper is organized as follows: Section \ref{sec: Background} offers an overview of the background and a survey of related work. Section \ref{sec: Methodology} elaborates on the methodology underlying \textbf{Mamba4Net}. Section \ref{sec: Experiment} details the experimental setups and presents the experimental results. Sections  \ref{sec:discussion} and \ref{sec: Conclusion} discuss and conclude the study, respectively.

\section{Background}\label{sec: Background}
\subsection{Large Language Models for Networking}\label{Bg:1}
Recent researches leverage LLMs4Net \cite{wu2024netllm} can be segregated into three primary paradigms: \textbf{zero-shot} adaptation, \textbf{fine-tuning} adaptation, and \textbf{reinforce-learning}(RL) integration. 

\textbf{Zero-shot} adaptation uses LLMs’ general-purpose knowledge directly, without additional training on task-specific data. This approach is widely applied in networking tasks such as anomaly detection \cite{jin2024large}, quality of service (QoS) analysis \cite{huang2024large} , and configuration management \cite{aykurt2024netllmbench}. For example, Sun et al. employed LLMs as aligners in UAV cluster networks to comprehend node status and issue high-level control directives \cite{sun2024large}. While this paradigm eliminates the overhead of specialized training, it may struggle to achieve the accuracy required in complex or highly specialized tasks.

In contrast, \textbf{fine-tuning} adaptation augments pre-trained LLMs with deeper, task-specific knowledge through training on domain-relevant datasets \cite{ding2023parameter}. This paradigm is particularly beneficial for high-precision or domain-specific tasks such as congestion control \cite{javaid2024leveraging} and traffic classification \cite{lavi2024fine}. It allows LLMs to internalize networking-focused information that builds upon (and refines) their existing language-based understanding.

Recently, there has also been increased interest in integrating \textbf{reinforce-learning} (RL) with LLMs, given the sequential decision-making nature of many networking problems. A notable example is NetLLM \cite{wu2024netllm}, which employs RL to fine-tune LLMs for cluster job scheduling (CJS) and adaptive bitrate streaming (ABR). By continually refining decisions based on feedback, RL-equipped LLMs can learn strategies tailored to specific network dynamics.

Despite these promising developments, LLMs4Net currently faces two major limitations. First, the attention mechanism in transformer-based LLMs typically exhibits quadratic time complexity \cite{hua2022transformer}, making it unsuitable for real-time processing of massive network traffic. Second, the large memory footprint of general-purpose LLMs hinders their deployment on resource-constrained devices, such as edge hardware. These issues pose a challenge not only in networking but also in natural language processing (NLP) more broadly. Although some recent studies have explored approaches to reduce time complexity \cite{xing2024spatial} or compress model scale \cite{zhu2024survey}, there is no existing method that simultaneously achieves a linearized attention mechanism and removes non-essential knowledge from LLMs. Addressing both of these goals is critical to enabling efficient, scalable, and real-time LLM-based solutions for next-generation networks.

\subsection{Motivation of leveraging Mamba backbone}\label{Bg:2}
The superior performance of current large language models is attributed to their Transformer backbone, whose self-attention mechanism is able to capture the global context \cite{soydaner2022attention}. While this mechanism is highly effective, it also leads to quadratic time complexity, which becomes a major bottleneck for large input sequences. Recent advances in selective state space sequence models (SSMs), particularly the Mamba architecture, demonstrate that similar global context capture can be achieved in linear time \cite{bicktransformers}. By leveraging state space representations to selectively model long-range dependencies, Mamba provides a more resource-efficient alternative to Transformers.

Credit for Mamba's linear complexity, it and its variants have been employed successfully across multiple domains. These domains including image processing \cite{zhuvision} and language modeling \cite{zhang2024mamba}. In these domains, Mamba has proven valuable for optimizing resource utilization \cite{zhou2024mdnet}. For instance, SegMamba \cite{xing2024segmamba} employs a Mamba-based model to segment 3D medical images, achieving competitive performance and overhead compared against Transformer-based approaches. These studies suggest that Mamba's linear complexity can lead to substantial gains in scenarios where computing resources and latency are critical concerns.

Mamba also have promising potential to be applied in networking domain. Differ from ordinary tasks, networking tasks running on resource-constrained devices often suffer from latency, bandwidth, and computational overhead, making the linear complexity of Mamba an attractive proposition. As an example, NetMamba \cite{wang2024netmamba} uses Mamba as an encoder to extract networking traffic representations for classification tasks, significantly reducing resource usage while maintaining robust performance. Nonetheless, most existing Mamba-based networking solutions still rely on the traditional deep learning paradigms, which require extensive engineering efforts for model development and deployment \cite{wu2024netllm}. This highlights a clear need for a generalized Mamba-based model that can streamline networking tasks and reduce overhead, enabling broader and more efficient application of linear-time architectures in networking operations.

\subsection{Knowledge Distillation of Large Language Models}\label{Bg:3}
Knowledge distillation is a common approach to compress LLMs by transferring task-relevant knowledge from a large teacher model to a smaller student \cite{yang2024survey, gu2024minillm}. Prior works mainly focus on distillation between Transformer-based models. To our knowledge, no study has addressed domain-specific cross-architecture distillation, highlighting the novelty of our research.

\section{Methodology}\label{sec: Methodology}
In this section we elaborate on the detailed design of Mamba4Net, a cross-architecture,  networking-specific knowledge distillation framework bridging LLMs with real-world networking application scenarios. Our goal is to support the two primary learning paradigms commonly adopted in networking - supervised learning (SL) and reinforce learning (RL) - similar to \cite{wu2024netllm}. Specifically, we implement Mamba4Net for three representative tasks: viewport prediction (VP) \cite{guan2019pano} \cite{han2020vivo} \cite{liu2023cav3}, adaptive bitrate streaming (ABR) \cite{kan2022improving} \cite{xia2022genet}, and cluster job scheduling (CJS) \cite{peng2021dl2}. 

\subsection{Overall Pipeline}\label{meth:1}
\begin{figure*}[htbp]
    \centering
    \includegraphics[width=\textwidth]{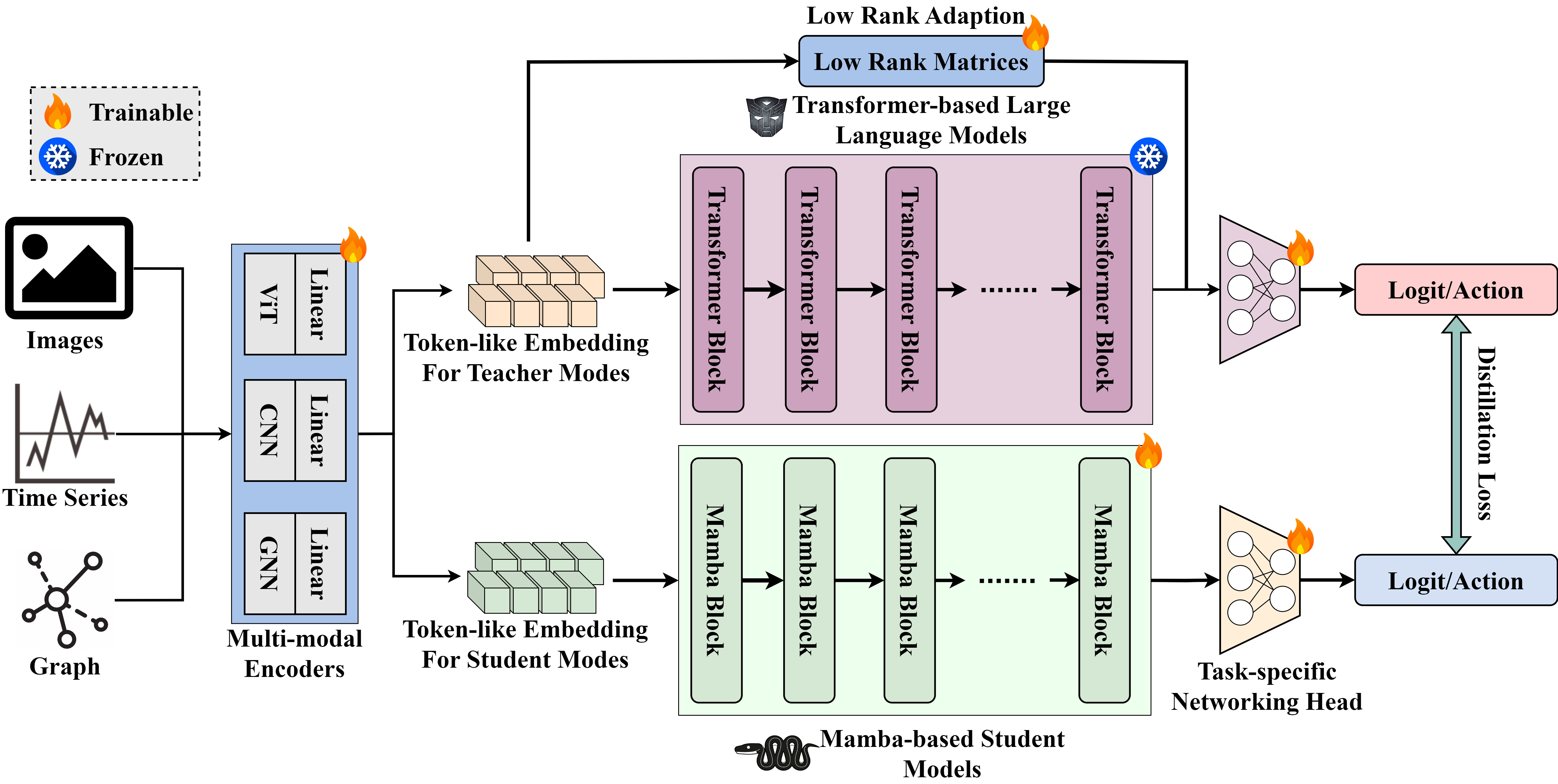} 
    \caption{Framework of Mamba4Net including Multi-modal Encoders, Task-specific Networking Head, and Teacher–Student Model with Cross-Heterogeneous Weight Reusing (CWR).}
    \label{fig:pipeline} 
\end{figure*}

As illustrated in Fig \ref{fig:pipeline}, the Mamba4Net pipeline comprises three primary components: multi-modal encoders, task-specific networking head, and teacher-student models' framework.

\subsubsection{Multi-modal Encoders}
Networking applications often involve diverse data modalities, such as video (for VP or ABR). Inspired by \cite{wu2024netllm}, multi-modal encoders in Mamba4Net are designed to process these heterogeneous inputs and convert them into unified token-like embeddings that LLMs can handle seamlessly.

Formally, each modality \(m\) has a dedicated encoder \(\mathcal{E}_m\), which maps the input \(x_m\) to a representation \(\mathcal{R}_m\):
\begin{equation}
    \mathcal{R}_{m} = \mathcal{E}_m(x_m).
\end{equation}

To save engineering cost, we leverage pre-existing mutual encoders specifically designed for each modality rather than constructing individual encoders for each modality from scratch. For instance, we employ the Visual Transformer (ViT) \cite{han2022survey} \cite{han2021transformer} to process image data for VP, and utilize Graph Neural Networks (GNN) \cite{wu2020comprehensive} to extract features from graph-based data for CJS tasks.

Notably, \(\mathcal{R}_m\) as the output of various encoders, its dimension often differs from the token size of LLMs. For instance, ViT-base typically outputs embeddings of dimension 768, whereas LLMs such as Llama accept token embeddings of size 4096. To address this, we employ a trainable linear layer:
\begin{equation}
    E = \text{concat}\left[\text{linear}(\mathcal{R}_m)\right]_m^M,
\end{equation}
where \(E\) denotes the final token-like embeddings concatenated across all modalities. The dimension of \(E\)can be adapted to different LLMs or to the Mamba-based student models. This step ensures that information of each modality is properly projected into unified embedding space, enabling seamless integration with LLMs and Mamba-based student models.

\subsubsection{Task-specific Networking Head}
 Each networking tasks rely on standardized output formats. For instance, output of VP is a probability distribution over possible viewports (e.g., top-k tiles or angles in a 360° video). For ABR, output provide a discrete set of bitrate decisions for each streaming segment. For CJS, output produce scheduling priorities or resource allocation decisions. However, default language model (LM) head of LLMs may generate invalid answer violating standardized output formats due to models' hallucination. 

To address this issue, a task-specific networking head is used in place of the default language model (LM) head. Custom networking heads are designed for different tasks, projecting the output logits of LLMs directly into the formatted answer, rather than relying on the LM to generate the answer, which alleviates the impact of models' hallucination.

\subsubsection{Teacher-Student Models' Framework}
Differing from prior work \cite{wu2024netllm}, Mamba4Net introduces a networking-specific knowledge distillation approach to reduce the computational overhead of adapting LLMs in real-world network applications. In this framework, the teacher model, denoted as \(\pi_t\), is a pre-trained transformer-based LLM, which provides networking-specific knowledge. The student model, \(\pi_s\), is a hybrid Mamba-based architecture that predominantly consists of Mamba blocks, with a limited number of transformer blocks near the output layer to maintain the stability of the distillation process. 

In structural level, there are two notable differences between the student model and teacher model. First, the number of layers in the student model is fewer than that of the teacher model. Second, the dimension of the token representation in the student model is smaller than that in the teacher model.

Based on the aforementioned characteristics of the student model, the size of the model can be significantly compressed while saving computational overhead. The detailed mechanism of cross-architecture distillation and cross-architecture weight reusing are illustrated in the following sections \ref{meth:2} and \ref{meth:3} respectively.

\subsection{Detailed Design of Cross-Architecture Networking-Specific Knowledge Distillation (DKO)}\label{meth:2}
In this section, we delve into the detailed design of the proposed DKO approach. The core ideas are twofold: (i) A knowledge distillation mechanism for both prediction tasks (supervised learning) and decision-making tasks (reinforcement learning), and (ii) Low-rank adaptation (LoRA) to efficiently infuse networking-specific knowledge into the teacher model, thereby constraining the fine-tuning process to a small number of parameters.

\subsubsection{Cross-Architecture Knowledge Distillation.} 
\subsubsection*{a) Supervised Learning (SL) Distillation}
Consider a supervised networking prediction task (e.g., viewport prediction (VP) \cite{chen2020sparkle}) with a dataset \(\mathcal{D} = \{(x,y)\}\), 
where \(x\) denotes the input features and \(y\) the label. 
Let \(\hat{y}_t\) and \(\hat{y}_s\) be the probability distributions (e.g., softmax of logits) output by the teacher and student models, respectively. 
The student’s training objective combines supervised loss with a distillation term:
\begin{equation}
\mathcal{L}_s = \mathcal{L}_{SL}(y, \hat{y}_s) + \alpha \, KL(\hat{y}_t \,\|\, \hat{y}_s),
\end{equation}
where $\hat{y}_t$ and $\hat{y}_s$ are teacher and student outputs. 
An optional temperature $\tau$ can be applied to smooth distributions.

\subsubsection*{b) Reinforcement Learning (RL) Distillation}
Another scenario for adapting LLMs to networking tasks involves decision-making via reinforcement learning (RL). Following the Mamba-based RL paradigm \cite{ota2024decision} \cite{rimon2024mamba}, we integrate policy distillation \cite{agarwal2024policy} to transfer knowledge from a transformer-based teacher model \(\pi_s\)

Similar to \cite{tamboli2024reinforced} \cite{janner2021offline}, we represent environment transitions as experience trajectories \(\mathcal{D}_{\text{rl}} = \{\tau_1,\dots,\tau_{|\mathcal{D}_{\text{rl}}|}\}\). 
Each trajectory \(\tau_n\) has \(T\) steps \(\{(r_t, s_t, a_t)\}_{t=1}^T\), where \(r_t\) is the reward, \(s_t\) is the state (environment), and \(a_t\) is the action. 
The cumulative reward is
\begin{equation}
    R \;=\; \sum_{t=1}^{T}\,r_t.
\end{equation}
Because networking actions/states often involve multiple dimensions\cite{wu2024netllm}, we decompose them into \(m\) components: 
\(a_t = \{a_t^1,\dots,a_t^m\}\) and \(s_t = \{s_t^1,\dots,s_t^m\}\). 
We then store the trajectory as
\begin{equation}
    \tau_t 
\;=\;
\bigl\{
   R,\,
   a_t^1,\dots,a_t^m,\,
   s_t^1,\dots,s_t^m
\bigr\}.
\end{equation}

We adopt an RL objective of the form
\begin{equation}
    \mathcal{L}_{\text{rl}}\;=\;\frac{1}{w}\sum_{t=1}^{T}\sum_{j=1}^{m}F_{\text{rl}}\!\bigl(a_t^j,\;\hat{a}_t^j\bigr),
\end{equation}
where \(\hat{a}_t^j\) is the student’s predicted action component, and \(F_{\text{rl}}\) encapsulates the reinforcement loss function (e.g., negative log-likelihood \(-\log\pi_s(a_t\mid s_t)\) times the reward or advantage), and \(w\) denotes a window or normalization factor (e.g., if we average over time steps). To encourage the student policy \(\pi_s\) to match the teacher policy \(\pi_t\), we add a distillation KL term:
\begin{equation}
\mathcal{L}_{\text{dis}}
\;=\;
\mathbb{E}_{s \sim \mathcal{D}}
\Bigl[
   KL\bigl(\pi_t(\cdot \mid s)\;\|\;\pi_s(\cdot \mid s)\bigr)
\Bigr].
\end{equation}
Here, \(\pi_t(a\mid s)\) and \(\pi_s(a\mid s)\) are probabilities over the action space. 
Our final RL objective then sums the RL loss and the distillation penalty:
\begin{equation}
    \mathcal{L}_{\text{RL-total}} 
\;=\;
\mathcal{L}_{\text{rl}}
\;+\;
\beta\,\mathcal{L}_{\text{dis}},
\end{equation}
where \(\beta\) is a weighting factor (analogous to \(\alpha\) in the SL case).

\subsubsection{Low-rank adaptation of teacher models}
Besides the knowledge distillation from teacher to student, the teacher model itself requires fine-tuning to acquire network-specific knowledge in the first place.Tuning all parameters of a large language model (LLM) is computationally expensive, so we employ a low-rank adaptation (LoRA) strategy. 
Given the teacher model's original weight matrix be \(\phi_{0}\in \mathbb{R}^{d_{out}\times d_{in}}\). We aim to find an update \(\phi_{\Delta}\) so that the final adapted weights are \(\phi=\phi_{0}+\phi_{\Delta}\). To reduce overhead, we freeze \(\phi_{0}\) and decompose \(\phi_{\Delta}\) into two low-rank matrices \(A\) and \(B\):
\begin{equation}
    \phi_{\Delta}=A\cdot \Sigma\cdot B,
\end{equation}
where \(A\in \mathbb{R}^{d_{in}\times r}\), \(B\in \mathbb{R}^{r\times d_{out}}\), and \(sigma\) is typically a diagonal or identity scale matrix (often omitted for simplicity). Consequently, the number of trainable parameters shrinks from \(d_{in}\times d_{out}\) to \(r\times(d_{in}+d_{out})\).

During forward propagation for an input \(x \in \mathbb{R}^{d_{{in}}}\), the teacher’s output is
\begin{equation}
\hat{y}
= \phi \cdot x
= 
\underbrace{\phi_0 \cdot x}_{\text{frozen}}
\;+\;
\underbrace{A\,\bigl(B \cdot x\bigr)}_{\text{low-rank update}}.
\quad 
\end{equation}
Because \(r \ll \min\bigl(d_{\text{out}},\,d_{\text{in}}\bigr)\) in practice, LoRA yields significantly fewer trainable parameters compared to full-model fine-tuning.

\subsection{Attention-to-Mamba Initialization}\label{meth:3}
In this section, we introduce the Cross-Heterogeneous Weight Reusing (CWR) mechanism, designed to reduce computational overhead and enhance robustness during knowledge distillation in Mamba4Net. Two key challenges motivate CWR: (1) the varying backbone architectures of teacher and student models, and (2) the mismatch in token sizes between them. To address these issues, CWR combines: (1) a cross-architecture weight reusing strategy that projects teacher weights onto the student’s structure, and (2) a low-rank approximation that compresses and regularizes transferred parameters. This approach preserves the teacher model’s valuable representations without incurring excessive computational cost, as the learnable projection aligns dimensions while minimizing distortion of the inherited features. Subsequently, the low-rank factorization further shrinks the parameter space, improving efficiency and mitigating overfitting. In practice, we observe that CWR not only accelerates the distillation process but also yields more robust performance against noisy inputs and distribution shifts. Detailed experimental results and ablation studies—highlighting the individual contributions of weight reusing and low-rank approximation—are provided in Section \label{sec: Introduction} to validate the effectiveness and stability of our proposed initialization method.

\subsubsection{Cross-Architecture Weight Reusing Strategy}
The reusing strategy (detailed in Algorithm \ref{alg:CWR}) initializes the Mamba-based student model with key weight components--namely \(Q,K,V\)--from a transformer-based teacher.Conceptually, this provides the student with a "linearized attention" warm start, leveraging the teacher's learned representations. 

\begin{algorithm}[!ht]
    \caption{Cross architecture initialization weight reusing}
    \label{alg:CWR}
    \begin{algorithmic}[1]     
        \STATE  \textbf{Shape:} \(B\)-Batch, \(L\)-Length, \(D\)-embed size, \(N\)-\(D\)/head,\ \(N'\) = expand heads,
        \STATE \textbf{Input:} \(o_t=(B,D)\)
        \STATE \textbf{Output:} \((B,D)\)
        \STATE \textbf{New Param:} \(\text{MLP},A\)
        \FOR{each head \(W^k_{lr},W^q_{lr},W^v_{lr},W^o_{lr}\):\((B,D)\) expanding grouped KVs}
        \STATE \textbf{Head Parameter}: \(A : (N, N')\)
            \FOR{for all positions $t$:}
                \STATE $x_t$: $(B, N) \leftarrow W^v_{lr} \otimes t$
                \STATE $B_t$: $(B, N) \leftarrow W^k_{lr} \otimes t$
                \STATE $C_t$: $(B, N') \leftarrow W^q_{lr} \otimes t$
                \STATE $\Delta_t$: $(B, N') \leftarrow MLP(x_t)$
            \ENDFOR
            \STATE $A_{1:T}$, $B_{1:T}$, $C_{1:T}$: $(B, N, N') \leftarrow \text{Disc}(A, B, C, \Delta)$
            \STATE $y \leftarrow \text{Mamba}(A, B, C, y)$
            \STATE output $\leftarrow \text{output} + W^T y$
        \ENDFOR
        
        \RETURN Outputs
    \end{algorithmic}
\end{algorithm}

\begin{figure}
    \centering
    \includegraphics[width=\linewidth]{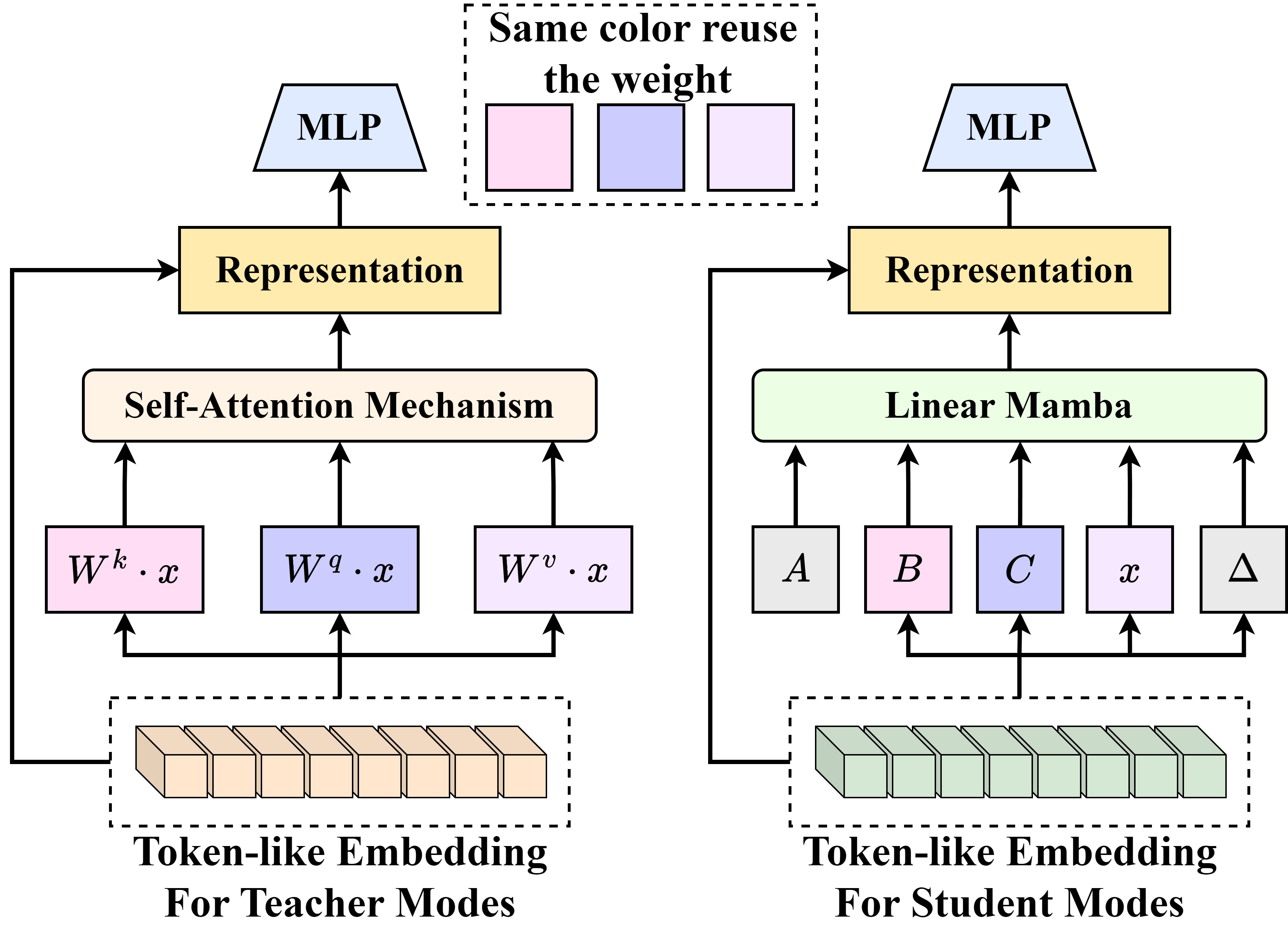}
    \caption{Structure diagram of Mamba and Transformer. In the diagram, there is a reuse relationship between weights of the same color in different architectures.}
    \label{fig:mamba}
\end{figure}

Fig. \ref{fig:mamba} illustrates how these parameters are mapped between a standard Transformer and our linear Mamba architecture, highlighting the structural differences and the process of adapting the teacher’s attention weights for the student.

\subsubsection{Low Rank Approximation for Dimension Mismatch}
A direct one-to-one transfer of weights can fail if the student and teacher have different embedding sizes. To address this, we use a truncated Singular Value Decomposition (SVD) to align dimensions while preserving the most influential components of the teacher’s weights.

Let the teacher’s weight matrix be \(\phi^T\in \mathbb{R}^{d_T\times N}\), where \(d_T\) is the teacher's embedding dimension and \(N\) is number of tokens. We perform a rank-\(r\) factorization:
\begin{equation}
    \phi^{(T)} = U_T \,\Sigma_T \,(V_T)^\top,
\end{equation} where \(U_T \in \mathbb{R}^{d_T \times r}\) contains the top \(r\) left singular vectors, \(\Sigma_T \in \mathbb{R}^{r \times r}\) is the diagonal matrix of the largest \(r\) singular values, \(V_T \in \mathbb{R}^{N \times r}\) contains the top \(r\) right singular vectors.

If the student's embedding dimension is $d_S$ (with $d_S \le d_T$), we construct 
\begin{equation}
    U_S \in \mathbb{R}^{d_S \times r},
\end{equation}by selecting the first $d_S$ rows of $U_T$. The initial student weight matrix 
\begin{equation}
    \phi^{(S)} \in \mathbb{R}^{d_S \times N}
\end{equation}is then given by:
\begin{equation}
\phi^{(S)} = U_S \,\Sigma_T\, (V_T)^\top.
\end{equation}

In this way, we reuse the most significant components of the teacher's weight matrix while matching the student's reduced dimension. This low-rank initialization both preserves critical information learned by the teacher and eases the computational load during training of the student model.

\section{Evaluation}\label{sec: Experiment}
\subsection{Experimental Setup}\label{sec: Experimental setups}
Based on previous LLMs4Net work \cite{wu2024netllm}, we set Llama2-7B as the foundation LLM of teacher model, and a hybrid Mamba-based model as student model. Then, we utilize Mamba4Net to conduct knowledge distillation in three various networking tasks VP \cite{wu2024mansy}, ABR \cite{xia2022genet}, and CJS \cite{mao2019learning}, detailed illustration is shown in Table. \ref{tab:networking_usecases}.

\begin{table*}[ht]
\centering
\caption{Information of three learning-based algorithm use cases in the networking area.}
\label{tab:networking_usecases}
\begin{tabular}{m{0.12\textwidth}m{0.12\textwidth}m{0.3\textwidth}m{0.15\textwidth}m{0.15\textwidth}c}
\toprule[1.5pt]
\textbf{Task} & \textbf{Paradigm} & \textbf{DNN Input} & \textbf{DNN Output} & \textbf{Objective}  \\ \toprule[1.5pt]
\textbf{Viewport Prediction} 
& Supervised Learning
& \textit{Time-series}: Historical viewports (eg: past viewing angles) ; \newline \textit{Image}: Video content information (eg: RGB; motion vectors)
& Predicted next frame or viewports 
& Minimize prediction error 
 \\ \hline
\textbf{Adaptive Bitrate Streaming} 
& Reinforcement Learning
& \textit{Time-series}: Historical network throughputs and delay; \newline \textit{Sequence}: Available chunk sizes at different bitrates; \newline \textit{Scalar}: Current buffer length 
& Optimal bitrate selection 
& Maximize user’s Quality of Experience and minimize rebuffering
\\ \hline
\textbf{Cluster Job Scheduling} 
& Reinforcement Learning
& \textit{Graph}: Dependency and resource demands of job execution stages described by DAGs
& Next job stage and number of executors allocated to the stage 
& Minimize job completion time \\ \toprule[1.5pt]
\end{tabular}
\end{table*}

We take NetLLM \cite{wu2024netllm}, the state-of-the-art LLMs4Net algorithm for comparison. Besides, we also take other learning-based approaches: TRACK \cite{rondon2021track} for VP, GENET \cite{xia2022genet} for ABR, and Decima \cite{mao2019learning} for CJS. Compared with NetLLM, there are two major differences beside backbone structure: (1) token embedding decrease from 4096 to 512, and (2) layer number reduce from 48 to 12 (includes 10 Mamba layers and 2 Transformer layers). The experiments are conducted on Linux server with two NVIDIA 40GB A100 GPUs.

\subsection{General Evaluation}
In this part, we conduct a comparative analysis of the general performance and computational overhead of Mamba4Net with those of other approaches to the three tasks previously mentioned in the testing environment. 

\begin{figure*}[htbp]
    \centering
    \includegraphics[width=\textwidth]{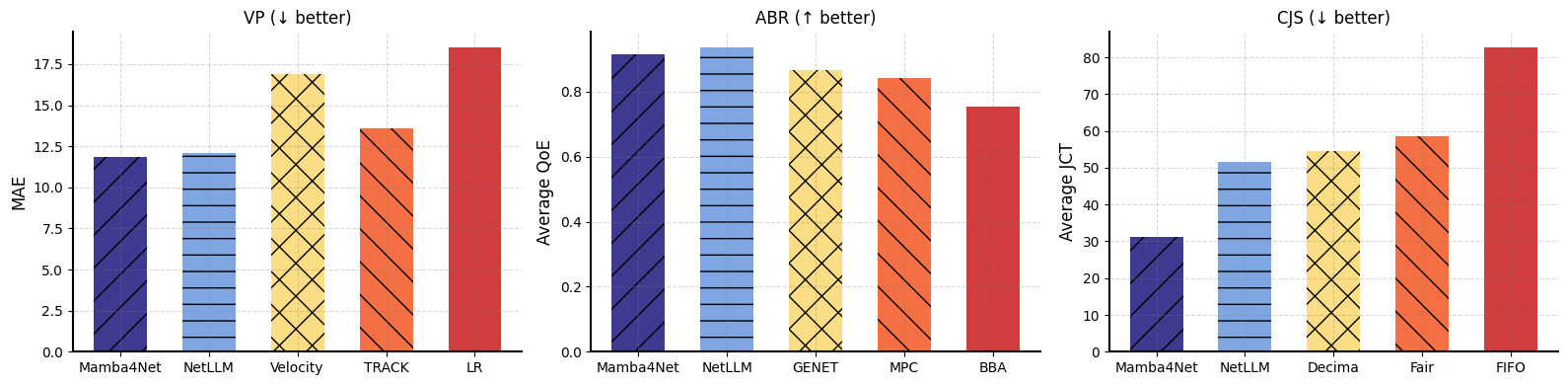} 
    \caption{Performance of Mamba4Net and its baselines in three various networking tasks.}
    \label{fig:exp1} 
\end{figure*}

Fig. \ref{fig:exp1} provides a comparative analysis of the general performance of Mamba4Net against multiple baseline methods in three different tasks. As demonstrated in Fig. \ref{fig:exp1}, Mamba4Net outperforms all baseline models in both the VP and CJS tasks. In the ABR task, Mamba4Net's performance closely approaches that of the state-of-the-art model, NetLLM, reducing 2\%-35.9\% in VP and 39.6\%-62.3\% of JCT for the CJS task. In the ABR task, Mamba4Net achieve only 2.4\% of QoE behind NetLLM. 

\begin{figure}
    \centering
    \includegraphics[width=\linewidth]{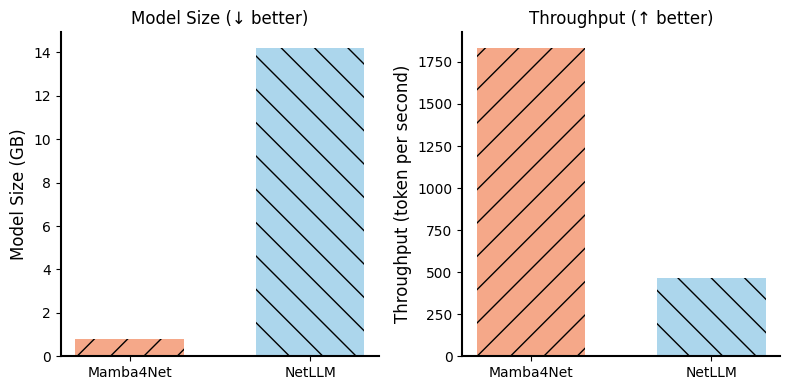}
    \caption{Model size and throughput of Mamba4Net compared against previous SOTA LLMs4Net approach NetLLM. In this evaluation, throughput only calculate the number of input token due to these special adaptions.}
    \label{fig:exp2}
\end{figure}

While the JCT metric, which is indicative of inference speed, underscores the efficacy of Mamba4Net, it does not demonstrate a substantial advantage over the previous state-of-the-art model, NetLLM, in terms of overall performance. However, Mamba4Net offers distinct advantages in terms of model size and throughput, as illustrated in Fig \ref{fig:exp2}, which compares the model size and throughput between Mamba4Net and NetLLM. Specifically, Mamba4Net attains 3.96 times the throughput of NetLLM while utilizing a mere 5.48\% of the model size.

According to the general evaluation, Mamba4Net model does not show a significant advantage over previous models in terms of accuracy metrics such as MAE and QoE. However, it exhibits a marked superiority in efficiency, particularly with regard to the JCT metric. This enhancement in efficiency can be attributed to the integration of the Mamba backbone, which operates with linear time complexity, thereby significantly enhancing the model’s throughput. Furthermore, by eliminating redundant knowledge, Mamba4Net achieves substantial model size compression, rendering its deployment more cost-effective and economically viable.

\subsection{Ablation Study}
In order to enhance comprehension of the contributions of distinct components within the Mamba4Net framework, an ablation study is conducted. The objective of this study is twofold: (1) to ascertain whether the knowledge distilled from LLMs contributes to the enhancement of Mamba4Net's performance, and (2)to determine whether Cross-Heterogeneous Weight Reusing (CWR) can substantially reduce the training time during knowledge distillation.

\begin{figure}
    \centering
    \includegraphics[width=\linewidth]{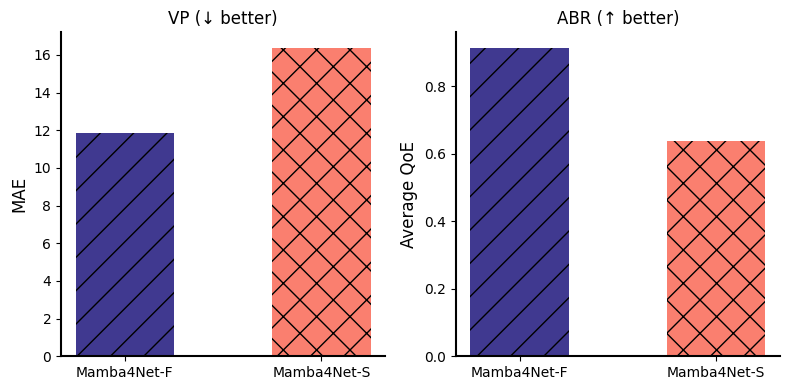}
    \caption{Ablation study on pre-trained knowledge of LLMs. \textbf{Mamba4Net-F} denotes model with LLMs as teacher model, \textbf{Mamba4Net-S} denotes student model trained directly in networking tasks.}
    \label{fig:exp3}
\end{figure}

\noindent\textbf{Impact of embedded knowledge of LLMs.} To gain a deeper understanding of how the pre-trained knowledge from LLMs impacts the student model, we design an ablation study focusing on the teacher model (LLMs). In this experiment, we remove the teacher model from Mamba4Net and train the student model directly on networking data, without the benefit of pre-trained knowledge from the LLM. In this experiment, we ignore CJS due to its evaluation metrics that focus on efficiency but not accuracy. Fig \ref{fig:exp3} demonstrates the results. In the VP task,  Mamba4Net with LLMs as its teacher model (Mamba4Net-F) outperforms Mamba4Net without a teacher model (Mamba4Net-S) by 38.1\% in terms of MAE; in the ABR task, Mamba4Net-F is 30.1\% ahead of Mamba4Net-S in term of average QoE.

The aforementioned experimental results demonstrate that pre-trained knowledge from teacher model (LLMs) has a substantial impact on the performance of Mamba4Net.

\begin{figure}
    \centering
    \includegraphics[width=\linewidth]{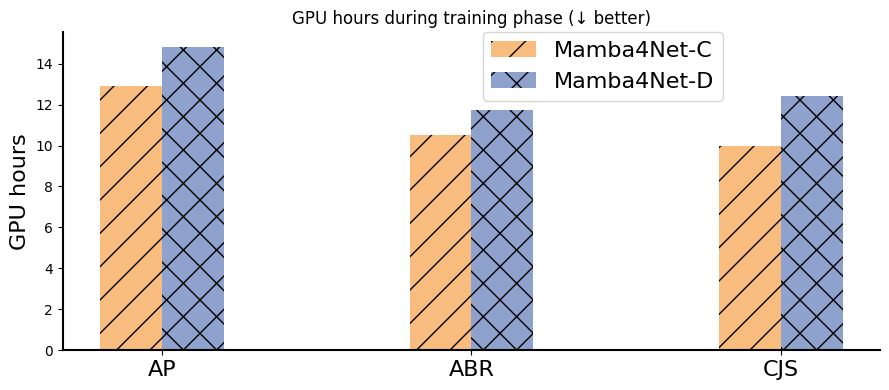}
    \caption{Ablation study on Cross Architecture Weight Reusing (CWR). \textbf{Mamba4Net-D} stand for student model initialized randomly and \textbf{Mamba4Net} stand for student model initialized through CWR.}
    \label{fig:exp4}
\end{figure}

\noindent\textbf{Impact of Cross Architecture Weight Reusing (CWR).} To verify the contribution of CWR in improving the efficiency of cross-architecture knowledge distillation, we removed CWR from Mamba4Net to create a control group (denoted as Mamba4Net-D) and compared it with the original Mamba4Net model that includes CWR (denoted as Mamba4Net-C). The findings of this study, as illustrated in Fig. \ref{fig:exp4}, demonstrates that CWR can reduce GPU hours by 10.31\%-19.77\% across diverse networking tasks.

This reduction can be attributed to CWR's utilization of a low-rank decomposition of LLMs weights during the initialization of the student model's weights. This process preserves a substantial amount of knowledge from the LLMs, thereby reducing the instability during the early stages of training. In traditional training, the student model must often learn a vast amount of knowledge from the beginning. However, CWR’s low-rank initialization strategy enables the student model to start training from a more stable point, thereby minimizing the risk of overfitting and accelerating convergence. 

\section{Discussion}\label{sec:discussion}
In this section, we discuss two critical aspects of the Mamba4Net framework: the significance of weight reusing in the distillation process and the importance of leveraging knowledge from Large Language Models (LLMs). These elements are pivotal in achieving the efficiency and performance gains demonstrated by Mamba4Net in networking applications.

\noindent \textbf{Importance of Weight Reusing in Distillation.} One of the primary challenges in knowledge distillation, particularly across heterogeneous architectures, is the effective transfer of learned representations from the teacher to the student model. Traditional distillation methods often rely on random initialization of the student model, which can lead to prolonged training times and instability, especially when the architectures differ significantly, as is the case with Transformer and Mamba models. In Mamba4Net, the teacher model is a Transformer-based LLM, while the student is a hybrid Mamba-based architecture. This architectural disparity necessitates a strategic approach to initialization to ensure stable and efficient training.

The Cross-Heterogeneous Weight Reusing (CWR) method addresses this challenge by transforming the teacher's attention block weights into compact factors through low-rank decomposition. These factors are then used to initialize the parameters of the Mamba-based student model. This process effectively narrows the representation gap between the two architectures, facilitating more efficient knowledge transfer. By initializing the student model with weights derived from the teacher, Mamba4Net ensures that the student begins training with a foundational understanding of the data, akin to that of the teacher. This approach not only accelerates convergence during the distillation phase but also enhances training stability, as evidenced by the reduction in GPU hours by 10.31\% to 19.77\% across various networking tasks (as shown in Fig. 6). Furthermore, the low-rank decomposition aids in compressing the model, contributing to the overall efficiency of Mamba4Net.

\noindent \textbf{Significance of LLM Knowledge for Mamba4Net.} Large Language Models (LLMs) are pre-trained on extensive datasets, endowing them with a broad spectrum of knowledge and capabilities. However, for domain-specific applications such as networking, much of this knowledge may be extraneous, leading to inefficiencies in both computation and memory usage. Mamba4Net addresses this by distilling only the networking-specific knowledge from the LLM, thereby creating a student model that is both compact and highly specialized. This targeted distillation process filters out generic content, allowing the student model to focus solely on the relevant information for networking tasks.

By leveraging the powerful representations learned by the LLM and specializing them for networking applications, Mamba4Net achieves superior performance. The ablation study presented in Fig. 5 underscores the significance of this approach, demonstrating that Mamba4Net with LLM knowledge outperforms the variant without it by 38.1\% in viewport prediction and 30.1\% in adaptive bitrate streaming in terms of their respective metrics. Moreover, by focusing on domain-specific knowledge, Mamba4Net achieves substantial model compression, reducing the model size to just 5.48\% of that of traditional Transformer-based models, as illustrated in Fig 4. This compression, combined with the linear time complexity of the Mamba architecture, enables Mamba4Net to operate efficiently in resource-constrained environments, a crucial requirement for real-world networking applications.

\section{Conclusion}\label{sec: Conclusion}
In this work, we have introduced Mamba4Net, an innovative framework that redefines the deployment of large language models (LLMs) in networking applications by addressing the long-standing challenges of computational inefficiency and memory constraints inherent to Transformer-based architectures. At its core, Mamba4Net leverages the linear computational complexity of the Mamba architecture, coupled with our novel Domain Knowledge-Oriented Cross-Heterogeneous LLM Distillation Method (DKO) and Cross-Heterogeneous Weight Reusing (CWR) techniques, to achieve unprecedented efficiency without compromising performance. This framework not only delivers a 3.96 \(\times\) improvement in throughput but also reduces model storage requirements to a mere 5.48\% of traditional Transformer-based models, setting a new benchmark for resource-efficient AI in networking.

The success of Mamba4Net lies in its ability to dynamically distill and transfer domain-specific knowledge from Transformer-based LLMs to Mamba-based student models, while filtering out irrelevant generic information. This ensures that the distilled models retain the critical expertise required for networking tasks. Furthermore, our CWR method bridges the architectural gap between Transformer and Mamba models by reusing and transforming high-dimensional attention weights into compact, low-rank representations, enabling stable and efficient knowledge transfer.

Through extensive experimentation on three pivotal networking tasks—viewport prediction, adaptive bitrate streaming, and cluster job scheduling—we have demonstrated that Mamba4Net consistently outperforms state-of-the-art methods, particularly in resource-constrained scenarios. These results underscore the framework's ability to deliver superior performance at a fraction of the cost, making it a practical and scalable solution for real-world networking challenges.

In conclusion, Mamba4Net signifies a paradigm shift in optimizing Large Language Models (LLMs) for domain-specific challenges, demonstrating exceptional combined strength in efficiency, scalability, and performance. By reimagining the distillation process and introducing innovative techniques for cross-architecture knowledge transfer, this work not only advances the field of networking research but also provides a blueprint for leveraging LLMs in other resource-constrained domains. In the future, we hope the philosophy of Mamba4Net can be employed in some resource-limited tasks like QoS optimization for wearable networking devices \cite{wu2024gazefed} and UAV cluster scheduling \cite{lu2024online}. We open-source our implementation to foster further exploration and invite the research community to build upon this foundation, paving the way for a new era of efficient and accessible AI solutions.

\section*{Acknowledgment}
This work is supported in part by National Key Research and Development Program of China (2022YFB2702300).

\bibliographystyle{IEEEtran}
\bibliography{reference.bib}

\end{document}